# Spontaneous spin selectivity in chiral molecules at the interface

Kouta Kondou[1], Shinji Miwa[2*], and Daigo Miyajima[1,3]

1. Center for Emergent Matter Science (CEMS), RIKEN, Wako, Saitama 351-0198, Japan
2. The Institute for Solid State Physics, The University of Tokyo, Kashiwa Chiba 277-8581, Japan
3. School of Science and Engineering, The Chinese University of Hong Kong, Shenzhen, Guangdong 518172, China

*miwa@issp.u-tokyo.ac.jp

**ABSTRACT:**
Chirality-induced spin selectivity (CISS) has been extensively studied over the past two decades. While current-induced spin polarization in chiral molecules is widely recognized as the fundamental principle of the CISS, only a few studies have been reported on bias-current-free CISS, where there is no bias electric current in chiral molecules. Recent studies on the chirality-induced exchange bias and current-in-plane magnetoresistance (CIP-MR) effects using chiral molecule/ferromagnet bilayer systems indicate that chiral molecules at the interface possess thermally driven broken-time-reversal symmetry, which induces bias-current-free CISS, i.e. a spontaneous effective magnetic field in the system. In this paper, we briefly review CISS-related phenomena in terms of the symmetry and discuss the mechanism of bias-current-free CISS. We also discuss the possibility of the linear magnetoelectric effect of chiral molecules, which arises from the spin polarization at the edges of molecules with metallic contacts, and its potential impact on the observed CISS phenomena.

## 1. Introduction

Chiral system has been recognized as an important component in recent research on spintronics [1], [2]. For instance, the spin polarization of electrons passing through chiral molecules has been extensively studied over the last two decades. This property is referred to as chirality-induced spin selectivity (CISS). CISS phenomena have been studied using various experimental techniques, such as photoelectron spectroscopy [3], [4], current-perpendicular-to-plane magnetoresistance (CPP-MR) measurements with a ferromagnetic electrode [5], adsorption-induced magnetization switching of an ultrathin ferromagnet [6], and enantiomer separation using a ferromagnetic substrate [7]. While current-induced spin polarization in chiral molecules is widely recognized as a fundamental principle of the CISS, only a few studies have been reported on bias-current-free CISS, where there is no bias electric current in chiral molecules [8]–[12]. In this paper, after briefly reviewing the CISS-related phenomena in terms of symmetry, we discuss the mechanism of bias-current-free CISS based on the recent studies of exchange bias [8] and current-in-plane magnetoresistance (CIP-MR) [12] effects. Furthermore, we also discuss the possibility of a linear magnetoelectric effect of chiral molecules at the interface and its potential impact on the observed CISS phenomena.

## 2. Structural chirality and CISS-related phenomena

Structural chirality is a pseudo-scalar that changes its sign under mirror reflection. It has recently been defined as an electric toroidal monopole [13], [14]. True and false chiralities have been practically defined as translating and stationary spinning cones [15], and circularly polarized photons and spin-polarized electric currents are classified as truly chiral systems. Right- or left-handed helices are often used to express chiral systems. The helicity of the helix is called axial chirality. When treating a real system for experiments, we can often be able to define an axial direction in terms of symmetry. For example, when an electric current is applied, the axial direction is parallel to the current path. In a multilayer system, the axial direction is the symmetry axis, which is perpendicular to the film plane. Thus, once a particular axis is determined, a right- or left-handed helix can be used as a toy model for a chiral system. In this section, we will also use an image of the helix to discuss the structural chirality.

### 2.1 Bulk spin polarization

While passing through chiral systems, the conduction electron can gain orbital angular momentum. This is analogous to an electric current in a small helix. The curvature of the helix couples the conduction electron and the orbital angular momentum. After orbital polarization, the conduction electron can be spin-polarized by the spin-orbit interaction, as shown schematically in Figs. 1(a) and 1(b).

### 2.2 Spin accumulation at the edges

When an electric current is injected from an electrode connected to chiral system, spin angular momentum can accumulate at the edges, as shown in Figs. 1(c) and 1(d). This can be explained as follows.

Using a linear-response two-current model for electric transport, which has been used to explain non-equilibrium spin accumulation and the MR effect in the ferromagnet/nonmagnet/ferromagnet junction [16], [17], chirality-induced bulk spin polarization and accumulation at the edge are schematically shown in Fig. 1(e)-(h). Note that the chiral system in Fig. 1(e)-(h) is treated using a solid-state band picture. In this respect, there is a big difference between the system in Fig. 1 and real system of chiral molecules. Here, the chiral system is contacted with two nonmagnetic electrodes (NM), where the electrical conductivity for up-spin and down-spin electrons are the same ($\sigma_{NM,\uparrow} = \sigma_{NM,\downarrow}$). The gray (orange) and blue curves represent the chemical potential for up-spin ($\mu_\uparrow$) (down-spin ($\mu_\downarrow$)) electrons and spin accumulation ($\mu_\uparrow - \mu_\downarrow$), respectively. The dashed line represents the pseudo-Fermi level. For the chiral system with right-handed helicity (Figs. 1(e) and 1(f)), the electrical conductivity for up-spin electrons ($\sigma_\uparrow$) is larger (smaller) than that for down-spin electrons ($\sigma_\downarrow$) for upward (downward) electron transport. For the system with left-handed helicity (Figs. 1(g) and 1(h)), $\sigma_\downarrow$ is larger (smaller) than $\sigma_\uparrow$ for upward (downward) electron transport. The difference in conductance is attributed to the momentum relaxation time, which depends on both the polarity of the spin angular momentum and the bias electric current. The introduction of such a momentum relaxation time is a striking feature of Fig. 1(e)-(h). In ferromagnets, such a difference in conductance is attributed to the spin-dependent density of state. We assume that the spin diffusion length in the system is smaller than its length [18]. Here, the sign of the bulk spin polarization depends on both the chirality and the direction of electron transport.

When the chiral system is contacted with nonmagnetic electrodes, a finite spin accumulation appears near the interfaces. The appearance of the spin accumulation can be derived from boundary conditions for charge and spin conservation at interfaces [16]. Here, the size of the spin accumulation at the interfaces depends on the spin diffusion length and the electrical conductivity in the materials. When the electrical conductivity in electrodes is small, spin accumulation is suppressed [19]. For this reason, there is no spin accumulation at the edges in freestanding chiral system as shown in Figs. 1(a) and 1(b). Note that in contrast to the bulk spin polarization, the polarity of the spin accumulation at the edges is independent of the electron transport direction and depends only on the chirality (see Fig. 1(c)-(h)). Consistent with these properties, the spin accumulation at the edges could be detected under alternating current [20].

**2.3 CISS-related phenomena**

As discussed above, spin polarization could be generated in the chiral system under an electric current. As pointed out in many previous studies, there are some problems in fully explaining the CISS-related phenomena by using the spin polarization in chiral molecules. Nevertheless, current-induced bulk spin polarization is often used to explain the CISS phenomenon on photoelectrons through chiral molecules [3], [4] and the CPP-MR effect observed in chiral molecules with a ferromagnetic electrode [5]. Although the spin-orbit interaction in the organic molecule may be small, the spin polarization is believed to be generated by the spin-orbit interaction in the molecule [21]–[23] or in the electrode [24]. Current-induced bulk spin polarization has been quantitatively studied in inorganic chiral materials, such as $CrNb_3S_6$ [25], [26]. Spin accumulation at the edges of chiral molecules, induced by the displacement current via molecular adsorption [7], has been proposed to explain CISS phenomena in enantiomer separation using ferromagnetic substrates [7] and adsorption-induced magnetization switching of ultrathin ferromagnetic metals [6].

In this way, CISS-related phenomena should arise from a result of current-induced bulk spin polarization or spin accumulation at the edges. Intuitively, there is no magnetic effect on the chiral molecule when no bias electric current is applied. However, there are only a few studies on bias-current-free CISS. More precisely, the chirality-induced spontaneous effective magnetic field has been experimentally observed in a chiral molecule/ferromagnetic bilayer system without injecting a bias current into the molecule. For example, such an effective magnetic field has been confirmed as an exchange bias effect [8], a magnetization-dependent molecular orientation [9], a suppression of magnetization fluctuation [11], and a CIP-MR effect [12]. In addition to these experimental observations, a model calculation study showed that molecular vibrations induce charge redistribution in the chiral molecule and accompany spin polarization when the chiral molecule is placed at the interface [10]. The theoretical study [10] is closely related to the bias-current-free CISS and will be discussed in detail later.

### 3. Observation of the Bias-current-free CISS

#### 3.1 Chirality-induced exchange bias effect

To study bias-current-free CISS, we used chiral phthalocyanine (Pc) molecules ((*P*)-PbPc-DTBPh and (*M*)-PbPc-DTBPh), which have right- and left-handed helicities, respectively [8]. A multilayer consisting of V (30 nm)/Fe (0.7 nm)/(*P*)- or (*M*)-PbPc-DTBPh (0-0.4 nm)/MgO(2 nm)/AlO$_x$(5 nm) was fabricated on a MgO(001) substrate, as shown in Fig. 2(a). A molecular layer of PbPc-DTBPh has a nominal thickness of ~0.6 nm. The black curve in Fig. 2(b) shows the magnetization hysteresis loop of Fe without a chiral Pc at room temperature. The magnetization hysteresis loop was collected using the magneto-optical Kerr effect. A magnetic field was applied perpendicular to the film plane. Perpendicular magnetic anisotropy was found in the Fe layers regardless of the thickness of the chiral Pc, while the adsorption of the chiral Pc at the Fe/MgO interface slightly changed the coercive field, i.e., the switching field (blue curve in Fig. 2(b)). Similar perpendicular magnetic anisotropy of epitaxial-like Fe/Pc/MgO system has been reported in Fe/CoPc/MgO multilayer [27].

Figure 2(c) shows the center field of the hysteresis curve, which is defined as the average of the positive and negative coercive fields. The pink and blue curves are guides for the eye. The center field decreases slightly as the molecular thickness increases, regardless of the helicity. This decrease is attributed to the fact that the adsorption of the chiral Pc molecules reduces the hybridization of the atomic orbitals of Fe and O. Here, we can see that there is a small but significant difference in the center field between Fe hysteresis adsorbed with chiral Pc molecules with right-handed and left-handed helicities where the nominal thickness of the chiral Pc molecule is more than half the molecular thickness (~0.3 nm). Considering this observation as the conventional exchange bias effect in a ferromagnet/antiferromagnet system [28], the experimental results indicate a chirality-induced effective magnetic field in the chiral Pc molecule. The direction of the effective magnetic field was upward for (*P*)-PbPc-DTBPh and downward for (*M*)-PbPc-DTBPh, as shown in Fig. 2(c).

#### 3.2 Chirality-induced CIP-MR effect

Bias-current-free CISS using chiral Pc was also confirmed in another system. We used a multilayer consisting of Ni(5 nm)/ (*P*)- or (*M*)-PbPc-DTBPh (0.6 nm)/MgO(2 nm)/AlO$_x$(5 nm) fabricated on a thermally oxidized Si substrate, as shown in Fig. 3(a) [12]. The multilayer was patterned into a 5 × 100 μm$^2$ rectangular shape using photolithography and Ar ion milling. The resistance of the sample was measured by the conventional four-terminal method.

Figure 3(b) shows the results of typical CIP-MR measurements at room temperature. A Ni sample without chiral Pc served as the control sample. A magnetic field was applied perpendicular to the film plane. All the sample resistances shown in Fig. 3(b) indicate the typical anisotropic MR effect of the Ni layer, as the magnetization direction of the Ni layer gradually rotates from in-plane to out-of-plane with increasing perpendicular magnetic field. The MR values in the positive and negative magnetic fields (±0.8 T) were almost the same in the Ni single-layer sample. In the Ni/(*P*)-PbPc-DTBPh bilayer sample, the MR value at +0.8 T was smaller than that at −0.8 T. Moreover, in the Ni/(*M*)-PbPc-DTBPh bilayer sample, the tendency was reversed; the value at +0.8 T was larger than that at −0.8 T. Note that the MR ratio is independent of the electric current amplitude from −100 to +100 μA (not shown, see Fig. S1 in Ref. [12]), indicating that the origin of the chirality-induced MR is not the same as that of the unidirectional MR [29]. The polarity of the chirality-induced spin polarization (or an effective magnetic field) deduced from the CIP-MR effect is consistent with the chirality-induced effective magnetic field confirmed in the Fe/(*P*)- and (*M*)-PbPc-DTBPh/MgO systems shown in Fig. 2 [8].

Figure 3(c) shows the temperature dependence of the chirality-induced CIP-MR. The MR ratio was defined as the difference in the resistance between magnetic field applications of +1 T and −1 T. With decreasing temperature from 300 K, the MR ratio shows a gradual decrease and is considered to be almost zero below ~50 K. This demonstrates thermally driven spin polarization (or an effective magnetic field) in chiral molecules on the Ni layer.

## 4. Mechanism of bias-current-free CISS

### 4.1 Thermally driven broken-time-reversal symmetry of chiral molecules at the interface

Non-equilibrium spin accumulation at the edges of chiral molecules is expected under bias electric currents (see Figs. 1(c)-(h)). In this context, without electric current injection, there is no spin accumulation in chiral molecules. However, our experimental observations indicate that, even without bias electric current in the molecules, the spin polarization can be induced at the edges at a finite temperature when chiral molecules are connected to electrodes (Fig. 4(a)-(d)). Such thermally driven broken-time-reversal symmetry of chiral molecules at the interface is consistent with recent theoretical work using model calculations [10]. In Ref. [10], molecular vibrations induce charge redistribution in the chiral molecules. Consequently, spin polarization occurs at the edges of the chiral molecule. The spin polarization at the interface itself or the associated linear magnetoelectric effect explains the bias-current-free CISS, that is, the exchange bias and CIP-MR effects.

**4.2 Possibility of linear magnetoelectric effect in chiral molecules**

The linear magnetoelectric effect is a phenomenon in which an applied electric (magnetic) field induces magnetization (electric polarization) [30], [31]. The multipole tensor is used to discuss the behavior of the linear magnetoelectric effect. Interestingly, the spin structure in Figs. 1(c), 1(d), 4(c), and 4(d) is reproduced by the magnetic monopole (*a*) and the $z^2$ quadrupole ($q_{z^2}$), as shown in Fig. 4(e). Here, both magnetic multipoles exhibit a linear magnetoelectric response in the *z*-direction as the following multipole tensor [32]:

$$\begin{pmatrix} a - \frac{1}{2}q_{z^2} & 0 & 0 \\ 0 & a - \frac{1}{2}q_{z^2} & 0 \\ 0 & 0 & a + q_{z^2} \end{pmatrix} \quad (1)$$

Thus, chiral molecules with an antiparallel pair of spin polarization at the molecular edges can exhibit a linear magnetoelectric effect. The spin polarization can be induced by placing a chiral molecule at the interface with conductive materials (Figs. 4(c) and 4(d)) or by injecting an electric current into chiral molecules from electrodes (Figs. 1(c) and 1(d)). In this way, the chiral molecules at the interface and/or under the current injection from the electrodes can be considered as multiferroics.

In Ref. [33], the electrical polarization of chiral molecules on the Au-coated Ni substrate was controlled by changing the magnetization direction of the Ni substrate, which was confirmed using Kelvin probe force microscopy. The spin accumulation shown in Figs. 1(c) and 1(d) can be excited by injecting electric current through the conductive cantilever. It is natural to assume that the effective magnetic field from Ni magnetization generates electric polarization in the chiral molecule via a linear magnetoelectric effect. In Refs. [34] and [35], finite magnetization was generated by an applied electric field in a junction consisting of a chiral molecule and metal [34] (or semiconductor [35]). In the experiments, applying the gate voltage to the chiral molecule induced an anomalous Hall voltage in the system, indicating that the electric field induced finite spin polarization. Note that the anomalous Hall voltage was maintained even after the displacement current injection. Therefore, the appearance of the anomalous Hall voltage may be related to the linear magnetoelectric effect resulting from the thermally driven broken-time-reversal symmetry of the chiral molecules at the interface (Figs. 4(c) and 4(d)).

**4.3 Explanation of bias-current-free CISS using the linear magnetoelectric effect**

The bias-current-free CISS phenomena, that is, the exchange bias (Fig. 2) and CIP-MR (Fig. 3) effects, are explained as follows. The exchange-bias effect in Fig. 2, that is, the effective magnetic field, could result from a combination of the linear magnetoelectric effect and the voltage-controlled magnetic anisotropy [36], [37]. Due to the linear magnetoelectric effect of chiral molecules resulting from thermally driven broken-time-reversal symmetry, as shown in Fig. 5(a), the electric polarization in chiral molecules changes with the magnetization direction of the ferromagnetic layer (FM in Fig. 5(a), Fe in Fig. 2(a)). The change in electric polarization induces a perpendicular magnetic anisotropy energy change of the ferromagnet via charge doping [38] and/or redistribution [39] at the interface. The effective magnetic field is considered as the magnetic anisotropy change depending on the magnetization direction.

Finite charge transfer plays an important role in the CIP-MR effect shown in Fig. 3. Electron transfer from the chiral molecule (PbPc-DTBPh) to the ferromagnetic layer (Ni) was confirmed by X-ray photoemission spectroscopy [12]. Electron transfer is also suggested by the increased electrical resistivity of Ni/PbPc-DTBPh/MgO compared to that of Ni/MgO [12]. As shown in Fig. 5(b), since the chiral molecule has electric polarization due to charge transfer, the chiral molecule has spin polarization due to the linear magnetoelectric effect resulting from the thermally driven broken-time-reversal symmetry. The spin polarization in chiral molecules induces the CIP-MR effect via spin-dependent scattering.

## 5. Potential impact of the linear magnetoelectric effect on the observed CISS phenomena

Starting with Ref. [5], there have been many reports on the chirality-induced CPP-MR effect. In ferromagnet/chiral molecule junction, the electrical resistance depends on both magnetization direction of the ferromagnet and molecular chirality. The MR effect has been confirmed in several experimental systems. A representative system is a chiral molecule on a ferromagnetic substrate (e.g., Ni) measured with a conductive atomic force microscope with a nonmagnetic cantilever [5]. The ferromagnetic material does not have to be a substrate but a cantilever (e.g., CoCr) [40]. A similar CPP-MR effect was also observed in a junction device consisting of a chiral molecule and a ferromagnetic metal (e.g., Ni) [41]. Interestingly, a similar MR effect was also observed in a system of a ferromagnetic electrode decorated with a chiral molecule measured using an electrochemical cell [42].

The CPP-MR effect in ferromagnet/nonmagnet/ferromagnet junctions [43] originates from the interface resistance, which can be derived from a two-current linear response model [16], [17]. However, when the same model is applied to the ferromagnet/chiral molecule junction, the MR effect does not appear [44]. While spin injection from the ferromagnet to the chiral molecule generates spin accumulation and interface resistance, which is analogous to a ferromagnet/nonmagnet junction, spin accumulation in the chiral molecule induces additional voltage due to the inverse effect of current-induced spin polarization, namely, the inverse CISS effect. In this way, the MR effect caused by the interface resistance is canceled out in two terminal measurements. A recent study has pointed out that a finite MR effect from spin-dependent transport may appear in a nonlinear regime [45].

Spin-dependent tunneling [46] is often used to discuss the magnitude of CPP-MR, which is known as the Julliere model (Figs. 6(a) and 6(b)). In this model, the MR ratio was determined as $(R_{max}-R_{min})/R_{max} = 2P_1P_2/(1+P_1P_2)$ using the spin polarization of the materials near the Fermi level ($P_1$, $P_2$). Although the Julliere model is useful for discussing the CPP-MR effect in magnetic tunnel junctions [47], it is difficult to believe that the model fully explains the chirality-induced CPP-MR. This is because the observed MR ratio [5], [48] is too large to be reproduced by the Julliere model using the spin polarization of the conduction electron in the chiral molecule ($P_1$) and that of the ferromagnetic electrode ($P_2$). Ni ($P_2 \sim 11\%$ [49]) is often used as a ferromagnetic electrode. While spin-dependent tunneling can be enhanced by spin blockade [50] or by increased spin polarization from the spinterface (half-metallic property at the molecule/ferromagnet interface) [51], [52], these effects are notable at low temperatures [53], [54]. Thus, it may be important to find a mechanism to explain chirality-induced MR that is free from the mechanism of spin-dependent tunneling.

A possible explanation for this is the magnetostrictive deformation of chiral molecules. The aforementioned linear magnetoelectric effect in chiral molecules can occur with magnetostrictive deformation. In this case, the ferromagnetic electrode does not act as a spin polarizer of the conduction electron (Figs. 6(a) and 6(b)), but as a source of the effective magnetic field via the interlayer exchange coupling [55], [56] and/or stray fields [57] to induce the linear magnetoelectric effect in chiral molecules. Since the molecule/metal interface possesses a Schottky barrier, a small change in the barrier height resulting from the magnetostrictive deformation drastically changes the electrical resistance in the system, as shown in Fig. 6(c) and 6(d). However, further experimental studies are needed to confirm this mechanism of the chirality-induced CPP-MR effect.

Recently, using a toy model consisting of an RC circuit and a magnetochiral diode, the relationship between electrical polarization and chirality-induced CPP-MR has been discussed [58]. In Ref. [58], the chirality-induced CPP-MR is a high-order effect and results from the charge accumulation induced by the current at the interface between the chiral molecule and the ferromagnetic metal. Indeed, as shown in Figs. 6(c) and 6(d), the Schottky barrier height can be modulated by the linear magnetoelectric effect via electrical polarization at the interface with the ferromagnetic electrode, apart from the magnetostrictive deformation of the chiral molecule.

## 6. Conclusion

In this paper, we review recent studies on CISS-related phenomena in terms of the symmetry. Besides the conventional CISS-related phenomena under bias electric current, there are only a few studies on bias-current-free CISS. Recent studies on the chirality-induced exchange bias and CIP-MR effects indicate that chiral molecules possess thermally driven broken-time-reversal symmetry at the interface, namely, thermally driven spin polarization at the edges. From the symmetry argument, the spin polarization at the edges can induce the linear magnetoelectric effect of chiral molecules and explain the bias-current-free CISS phenomena. In fact, the spin polarization at the edges is also induced by the bias current in the chiral molecule using metallic contacts, and may give us an explanation of large CPP-MR in terms of the linear magnetoelectric effect with magnetostrictive deformation or electrical polarization.

We believe that these discussions motivate further studies in the research field of chiral molecular spintronics.

**Acknowledgements**

We thank S. Sakamoto, M. Shiga, T. Hatajiri, M. Kobayashi, and Y. Otani of the University of Tokyo and H. Inuzuka, A. Nihonyanagi, and F. Araoka of RIKEN for collaborative research and discussions. We also thank K. Kimura of the University of Tokyo, H. Yamamoto of IMS, Y. Togawa of Osaka Metropolitan University, and T. Takenobu of Nagoya University for discussions. Part of this work was supported by JSPS KAKENHI (Nos. JPS22K18320, JP22H00290, and JP22H04964), Spintronics Research Network of Japan (Spin-RNJ), and the MEXT Initiative to Establish Next-Generation Novel Integrated Circuits Centers (X-NICS) (No. JP011438).

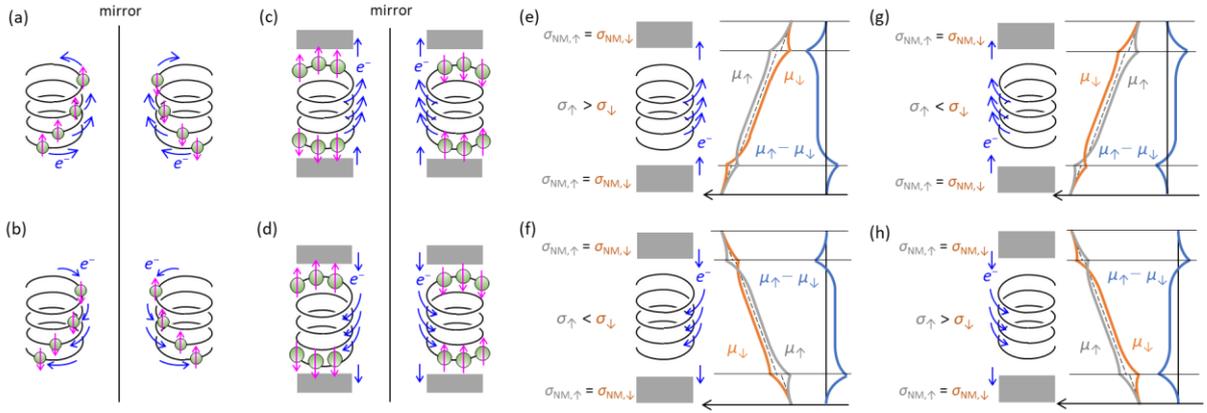

Fig. 1. (a), (b) Schematic of the current-induced bulk spin polarization. (c), (d) Schematic of the current-induced spin accumulation at edges when a chiral system is connected to electrodes (gray squares). The green circle, pink arrow, and blue arrow represents the electron, spin/orbital angular momentum, and electron flow, respectively. (e)-(h) Schematic of the electrochemical potentials. Chiral system is connected to two nonmagnetic electrodes (NM). The gray (orange) and blue curve represents the chemical potential for up-spin ($\mu_\uparrow$) (down-spin ($\mu_\downarrow$)) electrons and spin accumulation ($\mu_\uparrow - \mu_\downarrow$), respectively. The dashed line represents the pseudo-Fermi level. In the chiral system, the electrical conductivity for up-spin ($\sigma_\uparrow$) and down-spin ($\sigma_\downarrow$) electrons are different, and their relationship depends on both the chirality and the direction of the electron transport. In nonmagnetic electrodes, the electrical conductivity is the same for up-spin and down-spin electrons. The slope of each chemical potential is exaggerated. Note that the chiral system is treated using a solid-state band picture. In this respect, there is a big difference between the system in this figure and real systems of chiral molecules.

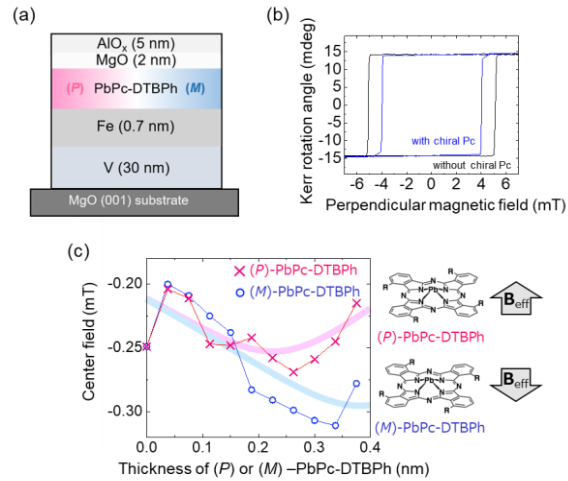

Fig. 2. (a) Schematic of the multilayer to observe the chirality-induced exchange bias effect. (b) Magnetization hysteresis curves of Fe with (blue, Fe/(*P*)-PbPc-DTBPh(0.3 nm)/MgO) and without (black, Fe/MgO) the chiral Pc adsorption at the Fe/MgO interface. (c) Center field of the hysteresis loop as the average of positive and negative coercive fields on the Fe decorated with (*P*)-PbPc-DTBPh and (*M*)-PbPc-DTBPh. Pink and blue curves are guides for the eye. Reproduced from [8] © 2020 IOP publishing on behalf of The Japan Society of Applied Physics (JSAP).

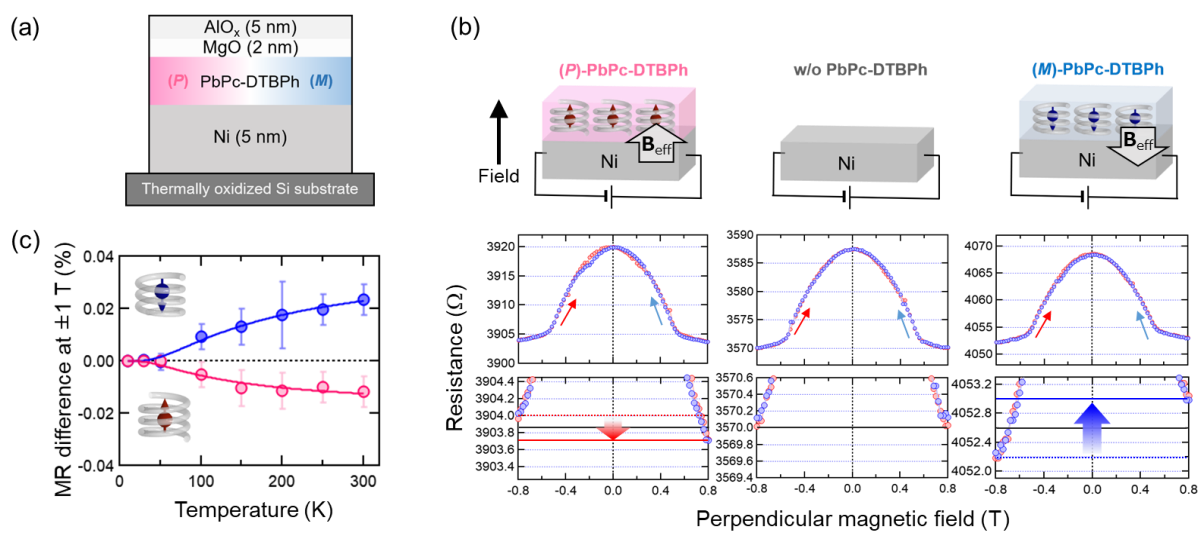

Fig. 3. (a) Schematic of the multilayer. (b) Chirality-induced CIP-MR effect. While the electrical resistance in the positive and negative magnetic fields (±0.8 T) was almost the same in the Ni single-layer sample, the resistance at +0.8 T was smaller (larger) than that at −0.8 T in the Ni/(*P*)-PbPc-DTBPh (Ni/(*M*)-PbPc-DTBPh) bilayer sample. (c) Temperature dependence of the MR effect. With decreasing temperature from 300 K, the MR ratio shows a gradual decrease and is considered to be almost zero below ~50 K. Reproduced from [12] © 2022 The Authors. Published by the American Chemical Society

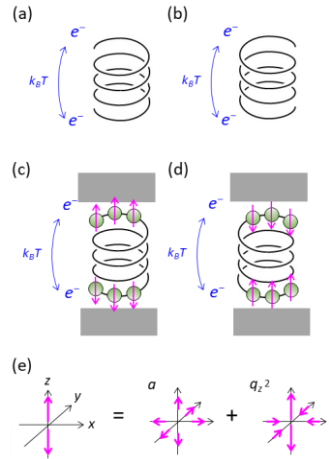

Fig. 4. Schematic of the freestanding and interfacial chiral systems at a finite temperature. (a), (b) While thermal fluctuation induces charge fluctuation, no spin polarization is induced in freestanding chiral systems. (c), (d) When a chiral system is connected to electrodes, finite spin polarization can be induced at the edges of chiral systems at a finite temperature. (e) Decomposition of the spin structure into magnetic multipoles. The green circle and pink arrow represents the electron and spin angular momentum, respectively.

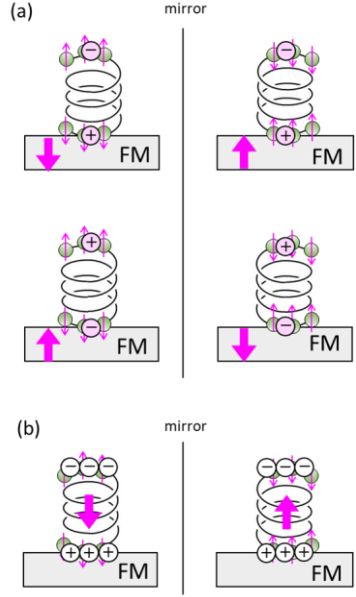

Fig. 5. (a) Schematic of the chirality-induced exchange bias effect. Due to the linear magnetoelectric effect resulting from the thermally driven broken-time-reversal symmetry, the chiral molecules have a chirality-dependent electric polarization (pink circles with + and −). The electric polarization changes the perpendicular magnetic anisotropy of the ferromagnetic layer (FM) via a voltage-controlled magnetic anisotropy effect. The pink arrow represents the spin angular momentum of the conduction electron or the FM. (b) Schematic of the chirality-induced CIP-MR effect. Due to the charge transfer between the chiral molecule and the FM (white circle with + and −), a finite spin polarization appears in the chiral molecule (pink arrows) due to the linear magnetoelectric effect. The sign of the spin polarization depends on the chirality and the electric polarization due to the charge transfer, independent of the magnetization direction of the FM.

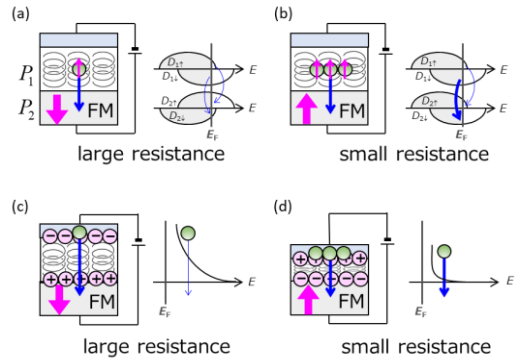

Fig. 6. (a), (b) Spin-dependent tunneling to explain the chirality-induced CPP-MR. Pink and blue arrow represents the spin angular momentum of the conduction electron in the chiral molecule or ferromagnetic electrode (FM) and the electron path, respectively. (c), (d) Magnetostrictive deformation and/or electrical polarization (pink circles with + and −) to explain the chirality-induced CPP-MR. The linear magnetoelectric effect in chiral molecules modulates the Schottky barrier at the molecule/FM interface and changes the electrical resistance in the system.